\documentclass[%
 aip,
 amsmath,amssymb,
reprint,%
]{revtex4-1}

\usepackage{graphicx}
\usepackage{dcolumn}
\usepackage{bm}
\usepackage{ulem} 
\usepackage[utf8]{inputenc}
\usepackage[T1]{fontenc}
\usepackage{mathptmx}
\usepackage{hyperref}
\usepackage{xcolor}

\begin{document}

\preprint{apl}

\title[WifiKID]{Contact-less phonon detection with massive cryogenic absorbers}
\author{J. Goupy}
\affiliation{Univ. Grenoble Alpes, CNRS, Grenoble INP, Institut N\'eel, 38000 Grenoble, France}
\author{J. Colas}
\affiliation{ENS Lyon, 15 parvis Ren\'e Descartes, 69342 Lyon, France}
\affiliation{Univ. de Lyon, Universit\'e Lyon 1, CNRS/IN2P3, IPN-Lyon, F-69622 Villeurbanne, France}
\author{M. Calvo}
\affiliation{Univ. Grenoble Alpes, CNRS, Grenoble INP, Institut N\'eel, 38000 Grenoble, France}
\author{J. Billard}
\affiliation{Univ. de Lyon, Universit\'e Lyon 1, CNRS/IN2P3, IPN-Lyon, F-69622 Villeurbanne, France}
\author{P. Camus}
\affiliation{Univ. Grenoble Alpes, CNRS, Grenoble INP, Institut N\'eel, 38000 Grenoble, France}
\author{R. Germond}
\affiliation{Department of Physics, Queen's University, Kingston, ON K7L 3N6, Canada}
\affiliation{Univ. Grenoble Alpes, CNRS, Grenoble INP, Institut N\'eel, 38000 Grenoble, France}
\author{A. Juillard}
\affiliation{Univ. de Lyon, Universit\'e Lyon 1, CNRS/IN2P3, IPN-Lyon, F-69622 Villeurbanne, France}
\author{L. Vagneron}
\affiliation{Univ. de Lyon, Universit\'e Lyon 1, CNRS/IN2P3, IPN-Lyon, F-69622 Villeurbanne, France}
\author{M. De Jesus}
\affiliation{Univ. de Lyon, Universit\'e Lyon 1, CNRS/IN2P3, IPN-Lyon, F-69622 Villeurbanne, France}
\author{F. Levy-Bertrand}%
\affiliation{Univ. Grenoble Alpes, CNRS, Grenoble INP, Institut N\'eel, 38000 Grenoble, France}
\author{A. Monfardini}%
 \email{alessandro.monfardini@neel.cnrs.fr.}
\affiliation{Univ. Grenoble Alpes, CNRS, Grenoble INP, Institut N\'eel, 38000 Grenoble, France}

\date{\today}

\begin{abstract}
We have developed a contact-less technique for the real time measurement of a-thermal (Cooper-pair breaking) phonons in an absorber held at sub-Kelvin temperatures. In particular, a thin-film aluminum superconducting resonator was realized on a $30$~g high-resistivity silicon crystal. The lumped-element resonator is inductively excited/read-out by a radio-frequency micro-strip feed-line deposited on another wafer; the sensor, a Kinetic Inductance Detector (KID), is read-out without any physical contact or wiring to the absorber. The resonator demonstrates excellent electrical properties, particularly in terms of its internal quality factor. The detection of alphas and gammas in the massive absorber is achieved, with an RMS energy resolution of about 1.4~keV, which is already interesting for particle physics applications. The resolution of this prototype detector is mainly limited by the low ($\approx 0.3$~\%) conversion efficiency of deposited energy to superconducting excitations (quasi-particles). The demonstrated technique can be further optimized, and used to produce large arrays of a-thermal phonon detectors, for use in rare events searches such as: dark matter direct detection, neutrino-less double beta decay, or coherent elastic neutrino-nucleus scattering.
\end{abstract}

\maketitle


Massive cryogenic detectors operated at sub-Kelvin temperatures are widely used in rare events searches, for example: the direct detection of dark matter\cite{EDW,CDMS, CRESST}, neutrinoless double beta decay \cite{CUORE, CALDER} searches, and quantitative studies of coherent elastic neutrino-nucleus scattering (CENNS) \cite{Billard:2016giu, Nucleus}. The current trend -in particular for dark matter and CENNS- points towards increased segmentation of the detector, i.e. more elements (crystals) and not necessarily more sensing elements for the same crystal/absorber, to provide the best trade-off between large target masses and low detection thresholds. 

Kinetic Inductance Detectors (KID) are thin-film superconducting resonators, sensitive to the content of superconducting excitations (quasi-particles) in the film. Variations of the kinetic inductance of the superconducting film causes the resonant frequency to shift around the nominal value; the change in kinetic inductance is caused by particle interactions breaking Cooper pairs, meaning the shift is proportional to the deposited energy. This detection principle, first proposed by the Caltech-JPL group \cite{Day2003}, has been integrated into arrays with thousands of pixels, used for example in millimeter wavelength astronomy\cite{Schlaerth2012, Adam2018} and single photon low-resolution spectrometers at visible to near-infrared wavelengths\cite{Meeker2018}. KID have also been used for single particle detection\cite{Swenson2010, Moore2012, Cardani2018}, and are a natural candidate for highly segmented detectors due to their high multiplex-ability.  

KID are capacitively or inductively coupled to the read-out/excitation line (feed-line), providing the unique possibility of realizing contact-less read-out lines. The advantage of this is twofold: first, the absorber can be prepared (or replaced) independently without any processing requiring wiring, and second, no thermal/electrical contact between the feed-line and absorber means that a potential loss mechanism for phonons is removed. The following describes the design, fabrication/packaging, and test setup of the contact-less KID detector, which consists of a superconducting resonator on a massive $30$~g silicon crystal absorber. The electrical performance of the resonator is characterized and compared to the design parameters, and the detector's single particle detection ability is demonstrated.

\begin{figure}[h]
    \centering
   \includegraphics[width=.4\textwidth]{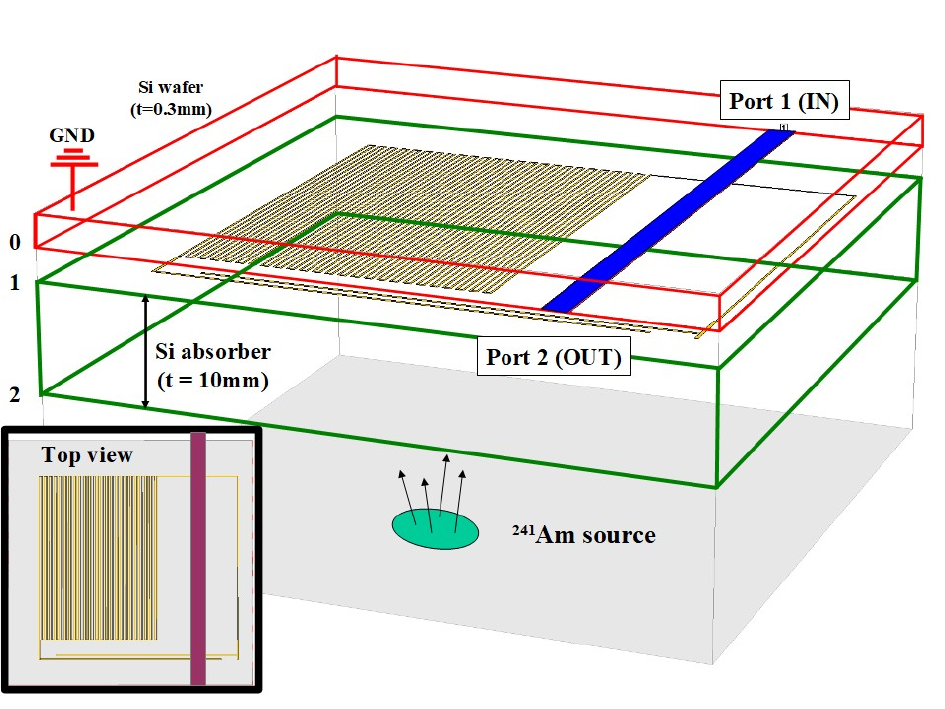}
    \caption{Three-dimensional schematic of the detector design. The $50~\Omega$ microstrip feed-line (blue) is realized on the (red) Si wafer (layer 0), and is separated from the 30~g Si absorber (green) by a $0.3~\textrm{mm}$ thick vacuum gap. The LEKID resonator is patterned onto the top face of the absorber (layer 1). A coupling loop is added between the meander and capacitor (right side of image), to minimize the impact of misalignment between the resonator and feed-line. Particles (alphas and gammas) from the radioactive source impinge on the bottom face of the absorber (layer 2). The thickness of the absorber has been reduced by a factor 10 for illustration purposes. Inset: top view.}
    \label{fig:design3D}
\end{figure}

A classical lumped element kinetic inductance detector (LEKID) design is used, see Fig.~\ref{fig:design3D}, based on a long ($\approx 230~\textrm{mm}$) and narrow ($20~\mu \textrm{m}$) inductive section, meandered to occupy a footprint of around $4 \times 4~\textrm{mm}^{2}$. Two capacitor fingers close the resonator circuit, which under the perfect lumped element approximation gives a resonance frequency of $f_r=(L\cdot C)^{-1/2}$. In this particular case, with only two capacitor fingers, the lumped element approximation is just close enough to estimate the order of magnitude of the geometrical inductance and capacitance: $L_{geom} \approx 110nH$, $C \approx 20pF$. In classical coplanar KID designs, the coupling is determined by the distance between the resonator and feed-line, which is precisely fixed by the lithography. For this contact-free detector, the coupling depends on the mechanical alignment and macroscopic distance between the feed-line wafer and the massive absorber. To reduce the sensitivity of the resonator's quality factor against possible misalignment, a coupling loop between the inductor and capacitor is added, by inserting an additional length of aluminum, with no meander, underneath the feed-line, see Fig.~\ref{fig:design3D}. The detector design was simulated using the Sonnet program (www.sonnetsoftware.com), and a detailed study of the effect of misalignment was performed.

\begin{figure}
    \centering
    \includegraphics[width=.35\textwidth]{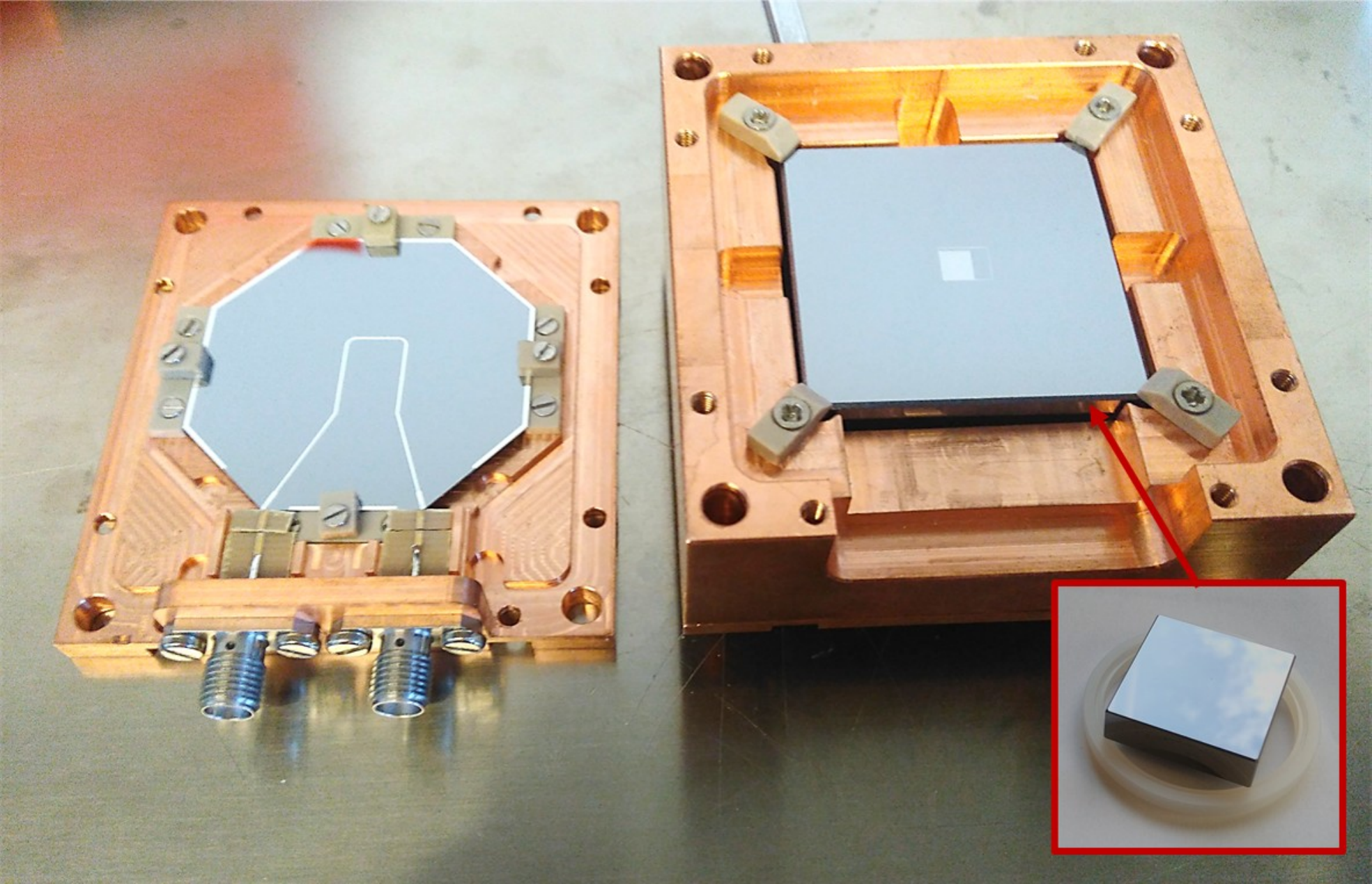}
    \caption{Picture of the two parts of the detector. On the left is the Si wafer with the superconducting micro-strip feed-line, wired to the IN/OUT connectors (ports 1 and 2). On the right is the resonator on the massive Si absorber, secured in the holder by PEEK clamps, which provide the thermal link.}
    \label{fig:holder}
\end{figure}

The absorber is a commercially available silicon crystal with dimensions of $36 \times 36 \times 10~\textrm{mm}^3$, a mass of roughly $30$~g and a resistivity exceeding $5~k\Omega \cdot \textrm{cm}$. The KID is realized on one of the two $36 \times 36~\textrm{mm}^{2}$ faces of the $\langle 100 \rangle$ crystal. The feed-line is realized on a separate, standard $300~\mu \textrm{m}$ thick silicon wafer. The metal is deposited by electron beam evaporation, under a residual vacuum of around $5\cdot10^{-8}~\textrm{mbar}$, at a rate of $0.25~\textrm{nm/s}$. Standard UV lithography ($\lambda=365~\textrm{nm}$) is then performed through a dedicated mask. The metal is patterned by a chemical step through resist apertures in a wet phosphoric acid bath. The fraction of the overall crystal surface that is covered by the metal is around 0.1\%. Two devices were produced, with nominal resonator film thicknesses of $20$~nm and $40$~nm respectively. See Fig.~\ref{fig:holder} for a picture of the detector. After considering, among other things, the natural aluminum oxidation, we estimate that the error in the residual thickness of the superconducting films is around $5$~nm for both cases. A copper detector holder was designed and fabricated. The absorber crystal is held by eight PEEK clamps, while four similar fixations are used for the feed-line wafer. The eight PEEK clamps represent the only thermal link $G$ between the massive silicon crystal and the cryostat. We estimate the thermal time constant $\tau_{th}$ of the system, at a temperature of $200$~mK, to be at least $20$~ms. 

The holder was mounted in a dilution refrigerator with a base temperature of $125$~mK. The underside of the Si crystal (opposite the KID) was irradiated with an $^{241}$Am source, collimated by a $6$~mm diameter hole in the holder. The KID signal is read-out with a homodyne system, in which a radio frequency (RF) synthesizer directly excites the KID at its resonant frequency \cite{Day2003}. The power reaching the KID is set to $-75$~dBm by a series of fixed (cryogenic) and variable (room-temperature) attenuators. A low noise ($T_{noise}\approx5$~K) SiGe HEMT amplifier (https://www.caltechmicrowave.org/amplifiers) is mounted on the $4$~K stage of the cryostat to amplify the output signal, which is then fed to a room temperature IQ mixer. This allows the \textit{inphase} (I) and \textit{quadrature} (Q) components of the output signal to be measured with respect to the input excitation. A fast digital oscilloscope reads the I and Q values, which are transmitted to an acquisition PC. These raw data pairs are then combined in order to extract time-ordered series of changes in resonance frequency (detuning parameter). This step allows removing nonlinearities.


The detector was cooled to the cryostat base temperature ($125$~mK). A heater, with PID control, is used to increase and stabilize the temperature, so studies of the resonator can be performed at different temperatures. The complex transmission ($S_{21}$) between ports 1 and 2 is determined from the I and Q measurements. At each temperature point, a calibration sweep is performed by sweeping the frequency of the RF synthesizer around $f_r$. The RF synthesizer is then set to the resonant frequency (determined by the calibration sweep), and streams of a few minutes are acquired, at a sampling rate of $1$~MHz. For ground-based astronomical applications, i.e. in the presence of strongly variable background, we routinely adopt a frequency-modulated readout technique to compensate in real time, among other things, the instantaneous quality factor variations\cite{Calvo2013}. This is not implemented yet in the fast readout used in the present study. We estimate that the calibration error introduced by this simplification is less than $10\%$, not affecting significantly the conclusions that will be presented. In the future it will be possible to further improve the energy calibration, for example by using multiple excitation tones.

\begin{figure}
    \centering
    \includegraphics[width=.5\textwidth]{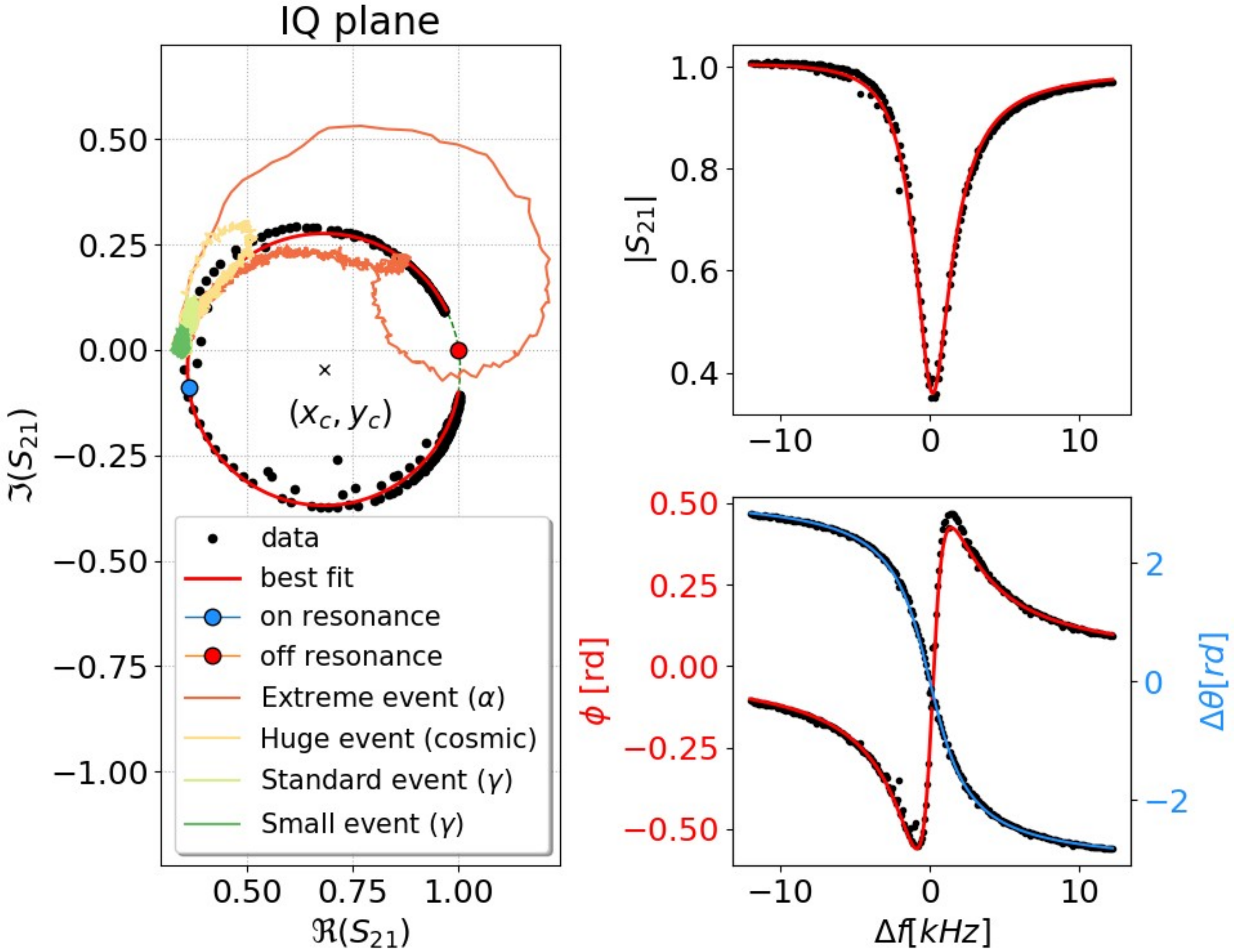}
    \caption{Example of the best fit of the complex transmission for the 20nm resonator after removing the influence of $a$ and $\varphi$ (see equation \eqref{eq:s21}). In this case, the base temperature was T$=200$~mK. The fitted quality factors are $Q_L = 1.49\cdot 10^5$, $Q_c = 2.22\cdot 10^5$, $Q_i = 4.50\cdot 10^5$ and the resonant frequency is $f_r= 564$~MHz. The phase $\phi$ is computed using the center of the complex plane as the reference and the phase variation $\Delta \theta$ is computed from the center of the calibration circle (black cross). Typical IQ trajectories of particle pulses are also shown. The systematic differences between the data and the fitted form are due to the approximate nature of equation \eqref{eq:s21}. }
    \label{fig:fit_probst}
\end{figure}

At base temperature the resonance was measured, as expected, at a frequency of $f_r=600$~MHz and $564$~MHz for the $40$~nm and $20$~nm devices respectively, see Fig.~\ref{fig:fit_probst}. To extract the resonator's electrical parameters, the measured $S_{21}$ is analysed with a standard procedure \cite{Probst2014}, and is fit to the following equation:

\begin{equation}
    \label{eq:s21}
    S_{21} = a e^{i\varphi}e^{-2\pi i f \tau}\left[ 1 - \frac{\left(Q_L/\vert Q_c \vert\right)e^{i\phi_0}}{1+2i Q_L\frac{\delta f}{f_0}}\right] = \text{I} + i\text{Q}.
\end{equation}
The parameters $a$ and $\varphi$ define an arbitrary affine transformation of the resonant circle. The impedance mismatch is in first approximation characterized by $\phi_0$, and $\tau$ is the cable delay. The internal ($Q_i$) and coupling ($Q_c$) quality factors of the resonator, even at $T=275$~mK, are both on the order of $2\cdot10^5$, resulting in a total (loaded) quality factor ($Q_L = 1/\left( Q_i^{-1}+ Q_c^{-1}\right)$) around $10^5$. At the lowest base temperature the resonator's $Q_i$ approaches $4\cdot10^5$. Similar quality factors were measured for both devices.

\begin{figure}
    \centering
    \includegraphics[width=.42\textwidth]{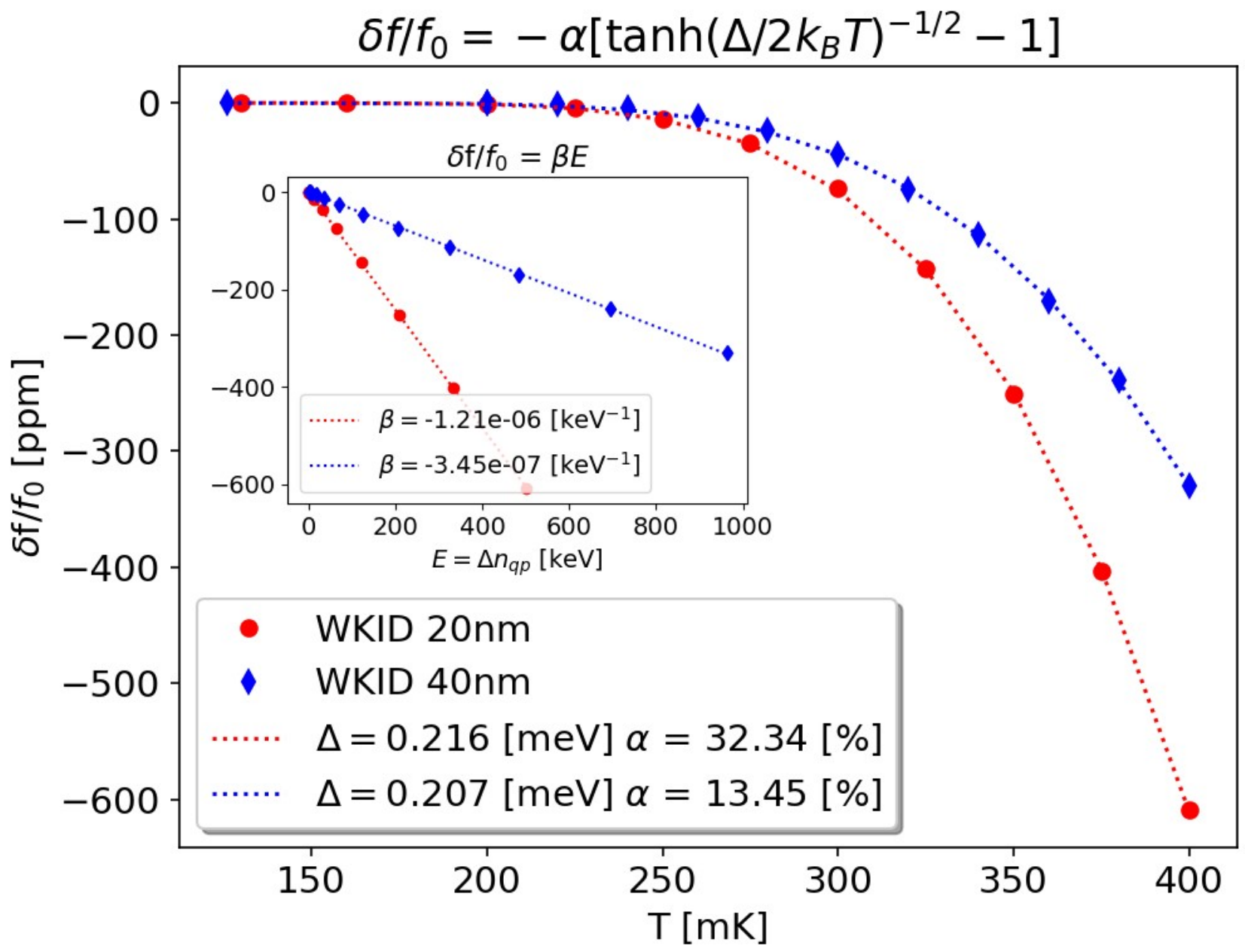}
    \caption{Dimensionless frequency shift data against temperature, along with a fit to equation \ref{eqn:MB_approx}. $\Delta$ and $\alpha$, the superconducting gap and the kinetic inductance fraction respectively, are free parameters determined by the fit. Inset (same y axis units): as predicted\cite{Swenson2010, Monfardini2014} the dimensionless frequency shift evolves linearly with respect to the number of quasi-particles $n_\text{qp}$.}
    \label{fig:ElecCharac}
\end{figure}

The detector responsivity, i.e the frequency shift as a function of the incident energy, is evaluated assuming that the response to thermal quasi-particles is equivalent to the response to quasi-particles generated by a-thermal phonons from the substrate~ \cite{GAO_nqp}. The change in resonance frequency as a function of temperature for the two devices is used to extract the kinetic inductance fraction ($\alpha$) and the superconducting gap of aluminum ($\Delta$) by performing a fit to an approximation of the Mattis-Bardeen theory \cite{Diener2012}, which holds for $hf \ll \Delta$ and $T < T_C/3$ (see Fig.~\ref{fig:ElecCharac}):
\begin{equation}
    \frac{\delta f}{f_0} = -\alpha \left[  \tanh{\left(\frac{\Delta}{2 k_B T}\right)} ^{-1/2} - 1 \right].
    \label{eqn:MB_approx}
\end{equation}
In this equation $f_0$ is the resonant frequency at $T \ll T_C$, and $\delta f = f_0(T)-f_0$ is the change in resonant frequency, as a function of temperature. The fitted value of the superconducting gap is then used to convert the temperature into energy with $E = N_{qp} \Delta$ and the following equation ~\cite{GAO_nqp} 
\begin{equation}
    N_{qp} = 2 N_0 V \sqrt{2 \pi k_B T \Delta} e^{-\Delta/k_B T},
\end{equation}
where $V$ is the volume of the resonator and $N_0$ is the single spin density of electron states at the Fermi energy. As shown in Fig.~\ref{fig:ElecCharac}, the frequency shift as a function of the energy follows a linear relation $\delta f/f_0 = \beta E$.  From the fit of this linear relation we estimated the detectors' responsivities $\beta$ (see insert of  Fig.~\ref{fig:ElecCharac}). 
In the limits $T \ll T_C$ and $h \cdot f \ll \Delta$ this linear relation is expected. The responsivity can also be estimated by the following formula:
\begin{equation}
    \beta=- \frac{\alpha}{2\cdot N_0 \cdot V \cdot \Delta^2}.
    \label{eqn:betaVSalpha}
\end{equation}
Numerical estimations, assuming\cite{GAO_nqp} $N_0 = 1.72 \cdot 10^{10} \mu m^{-3} \cdot eV^{-1}$, are $\beta(20nm)=-1.01 \cdot 10^{-6} keV^{-1}$ and $\beta(40nm)= -2.28 \cdot 10^{-7} keV^{-1}$.
The other parameters determined by the fit are consistent with their expected values, in particular we confirm that the superconducting gap increases when reducing the thickness, as expected for thin aluminum films \cite{PhysRevB.35.3188}.

The $20$~nm device was irradiated by an $^{241}$Am source with an $\alpha$ activity of $3$~kBq. The source produces $5.45$~MeV $\alpha$ and predominantly $60$~keV $\gamma$ particles, with respective rates of $100$~Hz and $30$~Hz in the crystal, according to Geant4 simulations~\cite{Allison:2016lfl} of the setup. 



\begin{figure}
    \centering
    \includegraphics[width=.4\textwidth]{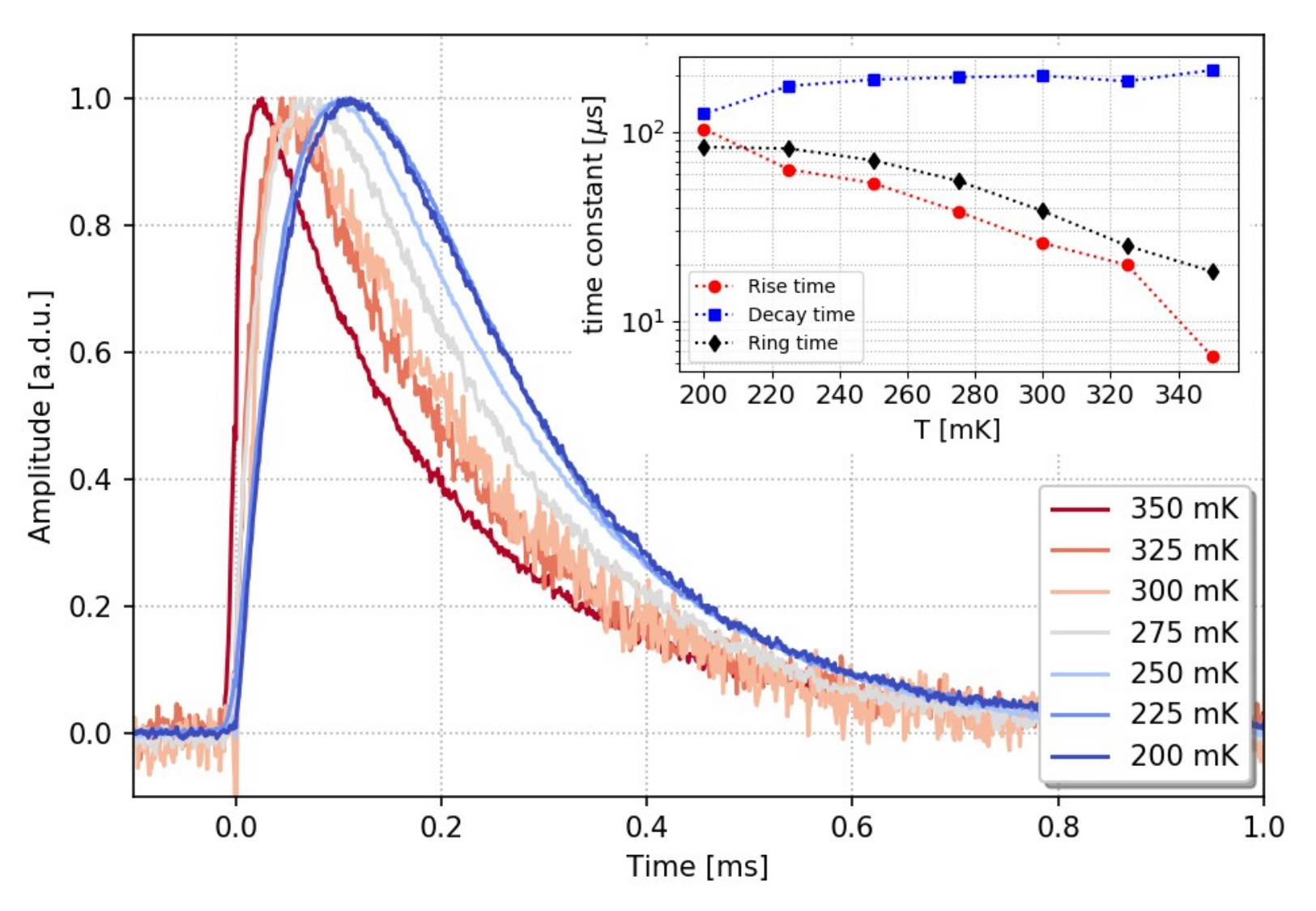}
    \caption{Examples of normalized pulses in the 20~nm thick device for different temperatures. Inset: evolution  with the temperature of the characteristics times of the exponential model (rise time: red circles, decay time: blue squares) compared to the calculated ring time of the resonator (black diamonds).}
    \label{fig:pulses}
\end{figure}

The pulse amplitudes, see Fig. \ref{fig:pulses}, are calculated by applying the optimal filtering technique \cite{Gatti1986}. Note that the conversion from phase to detuning is achieved using the calibration curve obtained after a circular fit of the $S_{21}$ data in the IQ plane, as shown in the right panel of Fig.~\ref{fig:fit_probst}. The calibration to detuning removes the non-linearities, in particular for large energy depositions, e.g. alpha particles. As the optimal filtering algorithm requires a pulse template, we chose an empirical model $f(t)$, based on our observations suggesting that only two time constants are relevant above $200$~mK. The analytical pulse template can then be written as  
\begin{equation}
\label{eq:model_pulse}
    f(t) = \Theta(t-t_0)\times \left(e^{-(t-t_0)/\tau_{\text{decay}}} - e^{-(t-t_0)/\tau_{\text{rise}}} \right),
\end{equation}
with $\Theta(t)$ the Heaviside function, $t_0$ the start time of the pulse, $\tau_\text{rise}$ and $\tau_\text{decay}$ respectively the rise and decay time constants of the pulse. These time constants were extracted by fitting multiple pulses simultaneously in the frequency domain. As shown in Fig.~\ref{fig:pulses}, we found that the rise time, along with the ring time of the resonator, decreases with the temperature for $T\geq$200~mK. The ring time is computed, based on the previously extracted total Q-factor values, as $\tau_\text{ring} = {Q_L(T)}/{\pi f_r(T)}$. The behaviours of the ring and rise times versus the temperature are well correlated in the range covered by the present study.
Contrary to the rise time, the decay time was found to be almost constant for temperatures up to $350$~mK, suggesting that the dominant relaxation process is related, at least above $200$~mK, to the lifetime of phonons in the absorber. It should be noted that for temperatures below 200~mK, we found some significant discrepancies between our simple two exponential pulse model, see Eq.~\eqref{eq:model_pulse}, and the data. Indeed, we have evidence that a third time constant is required to fully describe the pulse shapes. This additional time constant is most probably associated to the recombination rate of quasi-particles in the resonator film, which becomes comparable to the phonon lifetime at the lowest temperatures. A more precise study at $T<200$~mK is therefore required and will be undertaken in future publications. We remind that the thermal time constant of the detector has been estimated and lie in the order of the tens of milliseconds at $200$~mK. This is much longer than any time constant observed in our pulses, associated with Copper-pair breaking phonons. However, the exact influence of the PEEK clamps on the a-thermal phonons requires deeper investigations. According to a previous work~\cite{Martinez2019}, carried out on thin Silicon wafers and using classical KID, the losses through similar fixation points are non-negligible. In our case, having eliminated the feed-line and the bonding wires and working on a 3-D crystal, the relative weight of the fixation points losses versus the signal in the resonator might be even higher. 

\begin{figure}
    \centering
    \includegraphics[width=.4\textwidth]{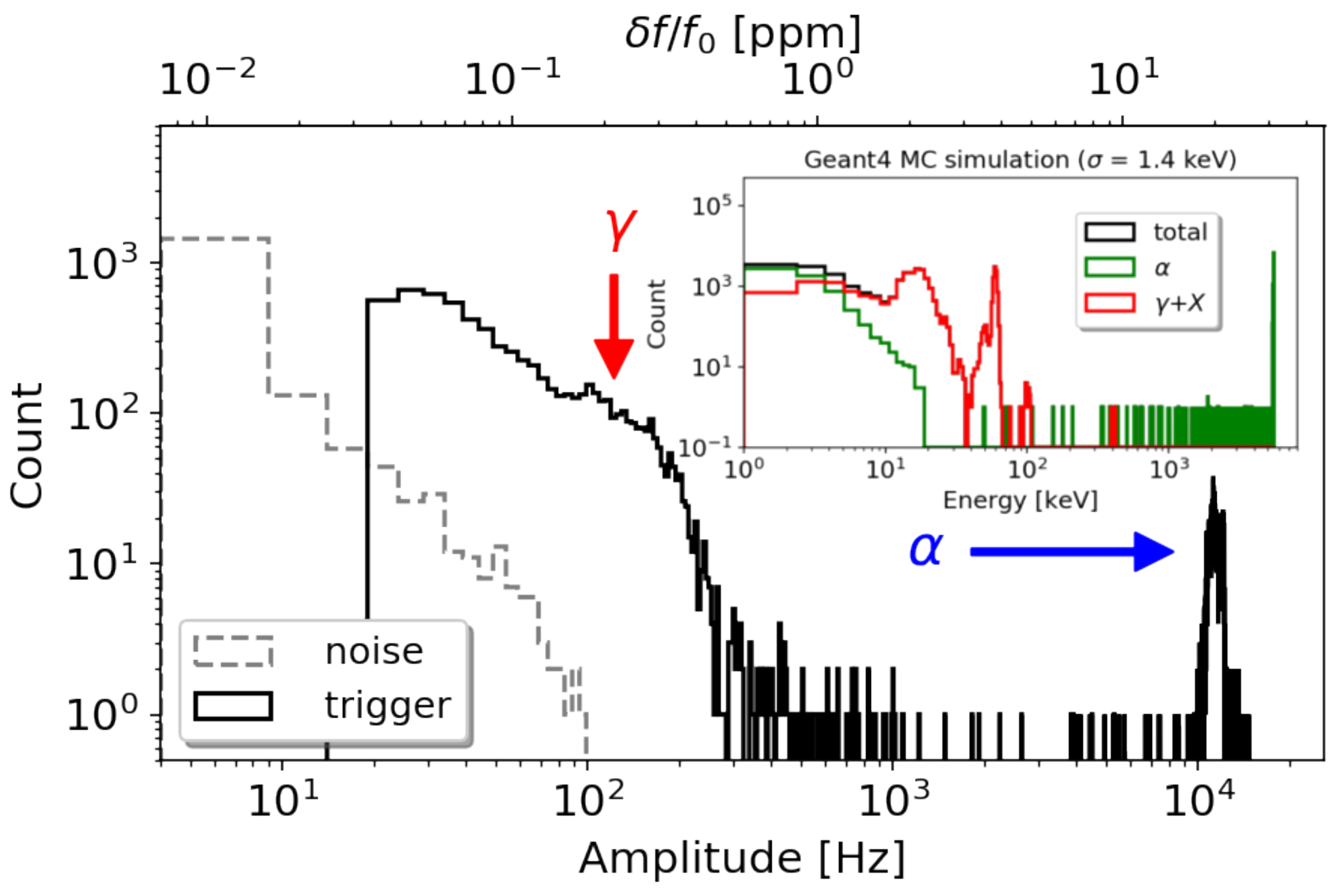}
    \caption{Distribution of fitted amplitudes for noise (dashed gray) and trigged events (solid black). Inset: comparison with Geant4 simulation. An energy resolution of 1.4~keV is considered, and the two populations of particles emitted by the Am source are taken into account, namely the $\alpha$ and $\gamma$. 
    }
    \label{fig:histo}
\end{figure}

The measured detuning (pulse amplitude) histogram is presented in Fig.~\ref{fig:histo} where the 5.45~MeV $\alpha$ peak is clearly visible at 11.2~kHz, along with a lower energy population with a bump around 100~Hz and an end-point at about 250~Hz. The baseline resolution in detuning units, estimated from the reconstructed amplitudes of noise samples, was found to be $\sigma_0 = $2.93~Hz (RMS). Using the energy calibration from the $\alpha$ peak, this then converts into a baseline energy resolution of $\sigma = $1.42~keV (RMS). It is worth noticing that with a keV-scale baseline energy resolution, one should expect to resolve the 60~keV line from the gammas, also emitted by the $^{241}$Am source, as suggested by the smeared Geant4 simulations (see inset panel of Fig.~\ref{fig:histo}). The fact that such a line, expected to peak around 123~Hz, is not resolved and that unlike the alpha particles these gammas interact almost uniformly between the two top and bottom surfaces of the crystal, strongly suggests that the measured pulse amplitudes depend on the location of the particle interaction inside the detector volume. The events lying near the 250~Hz end-point would represent, according to this interpretation, gammas interacting closer to the KID. 
Considering the 5.45~MeV alphas from the $^{241}$Am source impinging the Si crystal at the same location, opposite the KID, we can estimate the conversion efficiency of phonons to quasi-particles for this interaction to be $\eta \approx 0.3\%$. 


\begin{figure}
    \centering
    \includegraphics[width=.4\textwidth]{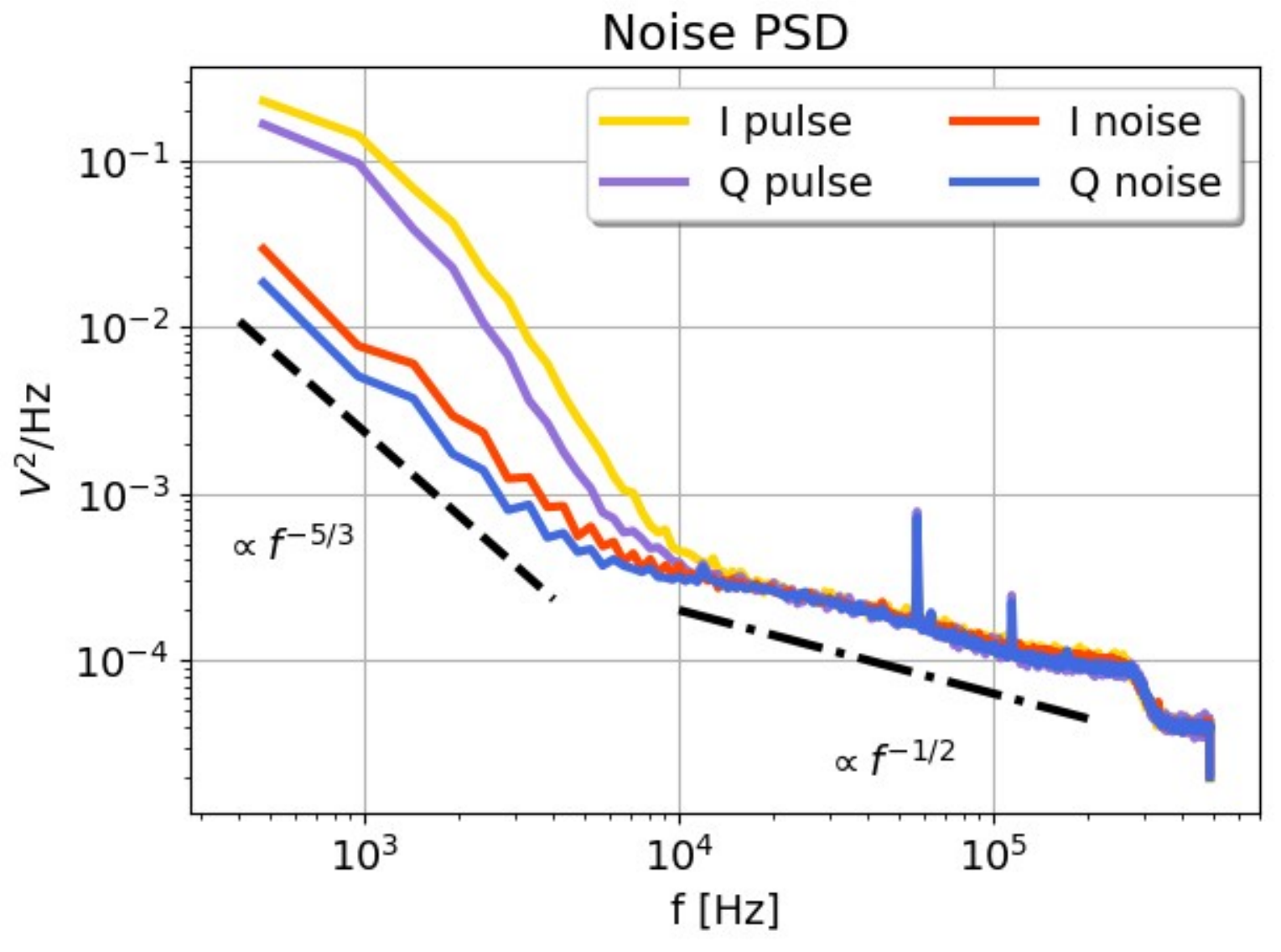}
    \caption{Noise and pulse Power Spectrum Densities measured at a temperature of $200$~mK for both I and Q components. The steeper $\propto f^{-5/3}$ component is attributed to vibration-induced sub-nanometer variations of the distance between the feed-line and the absorber crystal.}
    \label{fig:noise_psd}
\end{figure}
Finally, from our measured noise power spectral densities (see Fig.~\ref{fig:noise_psd}), we found, compared to many co-planar devices that we have tested in the past, an additional $1/f$-like component. We attribute a part of this excess noise to the relative variation of the distance between the feed-line and the resonator. Indeed, due to the resonator's sensitivity to changes in its electromagnetic environment, the KID is highly affected by changes in potential resulting from variations of the distance to the feed-line. Through electromagnetic simulations, we estimate that a variation of this distance of 1~\AA~translates into a shift of the resonance frequency on the order of $1\div2$~Hz. Therefore, in addition to increasing the KID sensitivity to energy deposition in the absorber, future efforts will be devoted to eliminating or reducing this specific environmental noise induced by vibrations. This could be achieved by running the experiment with a dedicated cryogenic suspension system to efficiently mitigate the vibration levels~\cite{Maisonobe_2018}. 



The main result of this work is the design and operation of a KID resonator with a contact-less feed-line that was deposited on a second wafer. We found that the resonator exhibits excellent electrical performance, suggesting that this sensor technology is well suited to be used in the context of particle detection. As well, this configuration is ideal to exclusively probe the contribution of the a-thermal phonons in detectors based on massive crystals. 
A first detector prototype, consisting of a single resonator implemented on a $30$~g high-resistivity Si crystal absorber has been tested, and an energy resolution $\sigma_E$ of about 1.42~keV (RMS) has been achieved, despite the low conversion efficiency of quasi-particles. These results are very encouraging in pursuing this development along the following lines: a) better understanding and modelling of the behaviour of a-thermal phonons in such massive crystals; b) optimizing the geometrical resonator design and fixations to increase the absorption efficiency by at least one order of magnitude; c) establishing a quieter test setup with lower vibrations to reduce the noise; d) using lower $T_C$ superconductors to increase the sensitivity to phonons (smaller gap) and the average quasi-particles recombination time.

Since~\cite{Cardani2015} $\sigma_E \propto \Delta^2 \cdot \tau_{qp}^{-0.5} \cdot \eta^{-1}$, the overall gain of the proposed improvements is estimated between one (prudent) to two (goal) orders of magnitude. 
It should be noted that a great advantage of this technology is that a large number of such devices can be connected in series thanks to a common read-out line, using external coaxial cables, naturally providing the required multiplexing for the future generation of highly-segmented detectors. Therefore, we believe that this technology could be used to build kg-scale detector payloads of $\mathcal{O}$(10)~g detector crystals. 

\begin{acknowledgments}
We are grateful, for inspiring discussions and help, to Andrea Catalano, Alain Benoit, Aurelien Bideaud, Marco Vignati and Angelo Cruciani.
\end{acknowledgments}


\bibliographystyle{aip}

\bibliography{WIKID_v2}

\end{document}